\documentstyle[sprocl]{article}

\input{psfig}
\def\Journal#1#2#3#4{{#1} {\bf #2}, #3 (#4)}

\def\PLB{{\em Phys. Lett.}  B}
\def\PRL{\em Phys. Rev. Lett.}
\def\PRD{{\em Phys. Rev.} D}
\def\ZPC{{\em Z. Phys.} C}

\def\ra{\rightarrow}

\def\be{\begin{equation}}
\def\ee{\end{equation}}
\def\bea{\begin{eqnarray}}
\def\eea{\end{eqnarray}}

\begin{document}

\title{PROSPECTS FOR MEASUREMENTS OF CP VIOLATION
AT HADRON COLLIDERS}

\author{ Jo\~{a}o P.\ Silva }

\address{Centro de F\'{\i}sica Nuclear da Univ. de Lisboa,
Av.\ Prof.\ Gama Pinto, 2\\
1699 Lisboa Codex, Portugal\\
and\\
Centro de F\'{\i}sica, ISEL}


\maketitle\abstracts{
A brief review of the strategies used in looking for CP violation in B decays
is presented. Some problems due to penguin diagrams are addressed.
Penguin trapping is discussed, in the context of the upcoming experiments.
The high rates and the possibility to perform experiments on the $B_s$
system justify the interest in hadronic colliders.}

\section{Introduction}

In the Standard Model, CP violation (CPV) arises from the clash of the Yukawa
couplings with the charged current interaction. When the Lagrangian is
written in the physical (mass) basis, this shows up as an irremovable
phase in the CKM~\cite{CKM} quark mixing matrix.
The magnitudes of most CKM matrix elements are constrained
experimentally~\cite{Neu96,reviews}, and exhibit an hierarchy,
with the magnitudes getting smaller as one leaves the diagonal.
Together with unitarity,
this motivates the Wolfenstein~\cite{Wol83} parametrization of the CKM matrix.
The Wolfenstein parameters $\lambda$ and $A$ are known to 1\% and
5\%, respectively. The best (correlated) constraints on the Wolfenstein
parameters $\rho$ and $\eta$ arise from the measurements of 
$|V_{ub}/V_{cb}|$~\cite{UkeBat}, $B^0 - \bar{B^0}$ mixing~\cite{GayCoy},
and the measurement of indirect CPV in neutral Kaons~\cite{Chr64}.

These constraints depend on hadronic matrix ``messy'' elements, and hence
have large errors. The corresponding allowed regions are shown in the figure,
taken with the gracious permission of Neubert from his recent 
review~\cite{Neu96}.
In this figure, the CP conserving measurements of $|V_{ub}/V_{cb}|$
and of $x_d = \Delta M / \Gamma$ (in the $B^0_d$ system)
already imply CP violation.
At present, this feature disappears if we take the most conservative bounds.


\begin{figure}
\centerline{\psfig{figure=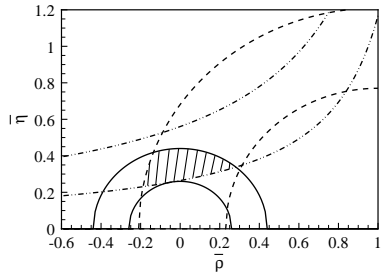,height=1.5in}}
\caption{Constraints on the $\rho$-$\eta$ plane. The circles centered
at (0,0) and (1,0) describe the bounds on  $|V_{ub}/V_{cb}|$
and $x_d$, respectively. The two hyperbolae arise from the CPV measurement
in neutral Kaons.
\label{fig:matthias}}
\end{figure}

Another way to express these constraints is to look at the orthogonality
of the first and third (db) columns of the CKM matrix. This can be written
as
\be 
1 = \frac{V_{ub}^\ast}{\lambda V_{cb}^\ast} + 
\frac{V_{td}}{ \lambda V_{cb}^\ast}.
\label{eq:triangle}
\ee
The squared magnitudes of the two terms on the RHS are
$\rho^2 + \eta^2$, and $(1-\rho)^2 + \eta^2$, respectively, 
and the corresponding bounds lead to the two circular sections in the figure.
Equation~\ref{eq:triangle} may be visualized as the well known 
unitary triangle in the $\rho - \eta$ plane.
It is constructed from a side starting at (0,0),
another at (1,0), and meeting at an apex lying within the
allowed region in the figure. Its angles are uninspiredly known as 
$\alpha$, $\beta$ and $\gamma$, going clockwise from the apex.
Increasing excitement is due to their upcoming measurement through
CP asymmetries in B decays.

Depending on how the $\epsilon^\prime/\epsilon$ experiments develop,
this might be the first independent measurement of CPV since 1964 (That
was several years before I was born!!). In any event, it will help to
constrain $\rho$ and $\eta$, shedding new light on the structure of the
CKM matrix and probing for New Physics effects.

\section{Searching for CPV in B decays}

Although CPV in the CKM matrix is a purely Electroweak phenomenon,
since quarks are included, we are unavoidably confronted with strong
interactions. These bring with them several new features:
\begin{itemize}
\item Though we can calculate reliably diagrams involving quarks and 
short-distance effects, the non-perturbative hadronic ``messy''
elements are not known to the desirable precision. This is responsible for
the large errors in the figure, and, unless such matrix elements cancel
out, also produces corresponding errors in B decays.
\item Besides tree diagrams, we also get gluonic-penguins, which will
complicate the analysis. There are also electroweak-penguins,
which play a crucial role in certain decays~\cite{Elwkpenguin}.
\item Strong phases are induced by final state interactions. Two such 
phases are mandatory for direct CPV, but may obscure the interpretation of
CPV from the interference between the mixing and the decay.
\item Since the strong interaction preserves flavour, one may rephase the
$B^0$ states. This means that only the clash between two phases may
have physical meaning.
\end{itemize}

The last point leads to three classes of asymmetry observables:
\begin{enumerate}
\item {\it Indirect CPV}, where we clash the phases of the 
$B^0 - \bar{B^0}$ mixing mass ($M_{12}$) and width ($\Gamma_{12}$).
We know that there is indirect CPV in the $K^0 - \bar{K^0}$ mixing,
due to the measurement of the $\mbox{Re}(\epsilon)$ parameter.
Unfortunately, this is expected to be unmeasurably small for the 
$B_d$ and $B_s$ systems.
\item {\it Direct CPV}, where we clash two independent decay paths.
This requires at least two amplitudes (preferably of similar magnitudes),
two different weak phases, and two different strong phases.
If the final state is not common to $B^0$ and $\bar{B^0}$, the decays
have the advantage of being self-tagging. 
This is most clearly seen in $B^\pm$ decays.
In the kaon system, a measurement of $\epsilon^\prime/\epsilon$ 
would be a signal of this type of CPV.
\item {\it Interference CPV}, occurring when the final state is common
to $B^0$ and $\bar{B^0}$, where one clashes the direct decay path,
$B \ra f$, with the indirect $B \ra \bar{B} \ra f$ decay.
\end{enumerate}
The first two methods are plagued by hadronic uncertainties. However,
these cancel in interference CPV when the (common) final state is a
CP eigenstate and the decay is overwhelmingly dominated by {\it one}
weak phase. One finds,
\be
a(t) = \frac{\Gamma[B^0(t) \ra f] - \Gamma[\bar{B^0}(t) \ra f]}{
\Gamma[B^0(t) \ra f] + \Gamma[\bar{B^0}(t) \ra f]} =
\zeta_f \sin(2 \Phi) \sin(\Delta M \, t),
\label{eq:asym}
\ee
where $\zeta_f$ is the CP eigenvalue of the final state $f$, and $\Phi$
is the weak phase. One can also perform a time-independent measurement,
thus paying a price of $x/(1+x^2)$. The measured value of $x_d$ yields a
modest suppression of $1/2$. However, in the $B_s$ system,
the current lower bound on $x_s$~\cite{GayCoy} already implies a
suppression of an order of magnitude. This is coming up on vertexing
limits~\cite{Quigg}, and makes time-dependent measurements mandatory
for the $B_s$ system.

The gold plated mode of this type is the $B \ra J/\psi K_S$ decay, which 
measures $\sin(2 \beta)$. This decay is important theoretically, because
the penguin and tree contributions share the weak phase, and also
experimentally, since the $J/\psi$ is easy to tag with di-leptons.
Current estimates for BABAR and BELLE predict uncertainties of order
$0.2$~\cite{FerAih}. At hadronic machines one has much higher luminosities,
but also a much harder environment to work on. Comprehensive Monte Carlo
simulations are still in their infancy. The fixed target HERA-B experiment
quotes uncertainties around $0.13$, while LHC-B quotes $0.023$~\cite{Ruc}.
In a very nice article, Butler~\cite{But95} obtains $0.07$,
for the Tevatron running at $10^{32}$ luminosity~\cite{Rij}.
I would like to thank him publicly for summarizing the effects of the
various cuts and efficiencies in a complete and clear table
(rather than scatter them all over 357 pages of jargon).

The hadronic mess reappears when one has more than one weak phase,
as it happens in $B \ra \pi^+ \pi^-$, where the tree and penguin contributions
come in with different phases. Here, besides the interference CPV
$\sin(\Delta M \, t)$ term, we also get a direct CPV $\cos(\Delta M \, t)$
dependence. Even if there were no final state phases
(in which case the $\cos$ term would vanish),
the penguin diagram would still affect the extraction of $\sin(2 \alpha)$
from the $B \ra \pi^+ \pi^-$ decay~\cite{Gro93}.
In fact, even for modest values of the penguin amplitude around
20\% of the tree amplitude, we can find deviations of up to $ \pm 0.4$ in
that determination\cite{Gro93}.

To remove this problem, Gronau and London~\cite{Gro90} have proposed
the use of isospin symmetry to relate this decay to $B^0 \ra \pi^0 \pi^0$
and $B^+ \ra \pi^+ \pi^0$. Though one must also take the electroweak
penguins into account~\cite{Des95}, these have been shown not to affect this
method~\cite{Elwkpenguin}.
Unfortunately, the  $B^0 \ra \pi^0 \pi^0$ branching
ratio is of order $10^{-6}$ and we would have to detect two neutral pions
in the final state. This is probably unfeasible at hadronic colliders.
Still, we can use $SU(3)$ symmetry and relate $B^0 \ra \pi^+ \pi^-$ to
the penguin dominated $B^0 \ra K^+ \pi^-$.
Silva and Wolfenstein introduced this idea to ``trap''
the penguin~\cite{Sil94}.
In fact, once $\beta$ is measured in 
$B^0 \ra J/\psi K_S$, all we need is the rate ratio~\cite{Sil94},
for which CLEO already has a (rather loose) bound~\cite{CLEO}.
Clearly, this technique will provide the first experimental handle on
$\alpha$. 
Recently, Gronau and Rosner~\cite{Gro96} have developed this idea,
showing that one may remove the penguin factorization hypothesis by
measuring in addition the branching ratio for $B^+ \ra K^0 \pi^+$.
This also allows for the independent determination of $\gamma$,
though $\beta$ will be measured beforehand anyway (recall that
$\alpha + \beta + \gamma = \pi$ in the SM; a fact used in all these
analysis).

Another approach consists in constructing polygons relating various
amplitudes to extract the CKM angles. For example, Gronau and 
Wyler~\cite{Gro91} have suggested the extraction of the angle $\gamma$ 
from triangles built using the $B^+ \ra K^+ (D^0,\bar{D^0},D^0_+)$ decays,
and their CP conjugates.
Subsequently, Dunietz applied this strategy to neutral decays\cite{Dun91},
in what the LHC-B collaboration calls Method Two~\cite{LHCB}.
For simplicity, I will use their notation.
Method One is due to Aleksan, Dunietz and Kayser~\cite{Ale92} and
uses $B_s$ decays into $D_s^\pm K^\mp$. 
One may also combine these polygon constructions with 
$SU(3)$ symmetry~\cite{Groetal} to extract the angles of the unitarity
triangle. Such techniques have been further explored in a very large number
of recent articles.

\section{The use of the $B_s$ system}

An important constraint on the parameters $\rho$ and $\eta$ would come
from a measurement of the mass difference between $B^0_s$ and 
$\bar{B^0_s}$, $\Delta m_s$.
Looking at figure~\ref{fig:matthias}, we realize immediately that the
constraint on $|V_{td}|$ arising from $\Delta m_d$ is rather poor.
This is due to the hadronic uncertainties :
the bag term and the decay constant for $B_d$ are only known to about 20\%.
However, their ratios to the corresponding factors for $B_s$ are much better
known. Therefore, a clean determination of $\Delta m_s / \Delta m_d$
would greatly improve our knowledge of the CKM matrix.
The current LEP limit~\cite{Coyle} is $\Delta m_s > 7.7 \mbox{ps}^{-1}$,
and the reader is encouraged to ask them about the meaning of the 
$8 \mbox{ps}^{-1}$ minimum in the likelihood plot.
Should this value become much higher, we might run into technology
limits~\cite{Quigg}.
Moreover, when looking for asymmetries in $B^0_s$ decays,
the oscillations in equation~\ref{eq:asym} will become too fast for
detection.

Fortunately, the width difference between $B^0_s$ and 
$\bar{B^0}_s$ might come to the rescue.
In a recent analysis, Beneke, Buchalla and Dunietz\cite{Ben96} find
\be
{\left(\frac{\Delta \Gamma}{\Gamma}\right)}_{B_s}
= 0.16^{+0.11}_{-0.09}.
\ee
Hence, the width difference cannot be neglected, and the decay rates
may be written as
\begin{equation}
\begin{array}{rcl}
\Gamma[B \ra f] & \propto &  
H(\Delta \Gamma\, t / 2) + I(\Delta M\, t)\\
& & \\
\Gamma[\bar{B} \ra f] & \propto &  
H(\Delta \Gamma\, t / 2) - I(\Delta M\, t).
\\ & &
\end{array}\label{eq:spa}
\end{equation}
For untagged decays, not only is there no statistics cost from tagging,
but the rapid oscillation $I(\Delta M\, t)$ terms drop out.
Strictly speaking, one should now talk about the short lived and the long
lived $B_s$ states.
As pointed out eloquently by Dunietz~\cite{Dun95},
this provides a uniquely different way to measure CPV in the B system.
In particular, Method One is applicable with untagged samples.
Dunietz~\cite{Dun95} also proposes an untagged $B_s$ into $D^0 \phi$
version of Method Two.
This is a hot topic, and rapid new developments are expected.

\section{Conclusions}

The fact that the SM has only one CPV independent quantity makes it 
very predictive, giving us an ideal setting in which to look for the
Physics Beyond. Since thirty two years is a lot of time to seat on our hands,
this picture should (and will) be tested.

In the B system, the first information will be the measurement of
$\beta$ in the gold plated $B \ra J/\Psi K_s$ decay.
For a measurement of $\alpha$ in $B \ra \pi^+ \pi^-$ we must trap the penguin.
This can be performed~\cite{Sil94} comparing it to $B \ra K^+ \pi^-$,
and we can avoid the penguin factorization approximation~\cite{Gro96}
by measuring in addition $B^+ \ra \pi^+ K^0$.
As for the extraction of $\gamma$, there are several classic 
methods~\cite{Gro91,Dun91,Ale92}.
The good news from Dunietz~\cite{Dun95} is that we may be able to use
untagged samples for Method One~\cite{Ale92},
and for the $B_s$ into $D^0 \phi$ version of Method Two~\cite{Dun95}.
This is specially interesting if the width and mass differences in the 
$B_s$ system turn out to be large.

The angle $\beta$ will soon be measured at BABAR, BELLE and HERA-B.
We will also learn a considerable amount about penguin trapping.
The high rates and the (complementary) $B_s$ Physics opportunities available
at the Tevatron and LHC warrant dedicated B-Physics experiments.
Understanding the real capabilities of such experiments, under what
is admittedly a very hard environment, will require extensive
Monte Carlo simulations.
Please summarize each step of your cuts and efficiencies clearly.
We will thank you for it.

\section*{Acknowledgments}

I would like to thank the Organizing Committee, and specially Prof.\ Luca
Stanco, for inviting me to Abano and for making this workshop so enjoyable.
I am extremely grateful to Prof.\ Matthias Neubert for giving me his
figure~\ref{fig:matthias}, and allowing me to reproduce it.
I am also greatly indebted to Prof.\ Wolfenstein for extensive discussions
and for reading this written version.
This work was supported by JNICT under project CERN/P/FAE/1050/95.



\section*{References}
\begin{thebibliography}{99}
\bibitem{CKM}N. Cabibbo, \Journal{\PRL}{10}{531}{1963};
M. Kobayashi and T. Maskawa, \Journal{\em Progr. Theor. Phys.}{49}{652}{1967}.

\bibitem{Neu96}M. Neubert in {\em hep-ph/9604412}.

\bibitem{reviews}For other recent reviews see, for example,
A. Pich and J. Prades, \Journal{\PLB}{346}{342}{1995};
A. Ali and D. London in {\em hep-ph/9508272};
A. Pich, in {\em hep-ph/9601202}.

\bibitem{Wol83}L. Wolfenstein, \Journal{\PRL}{51}{1945}{1983}.

\bibitem{UkeBat}F. Ukegawa, these proceedings;
M. Battaglia, these proceedings.

\bibitem{GayCoy}C. Gay, these proceedings;
P. Coyle, these proceedings.

\bibitem{Chr64}J.H. Christenson, J.W. Cronin, V.L. Fitch and R. Turlay,
\Journal{\PRL}{13}{138}{1964}.

\bibitem{Elwkpenguin}For recent reviews see, for example,
M. Gronau {\it et. al}, \Journal{\PRD}{52}{6374}{1995};
A.J. Buras, \Journal{\em Nucl. Instrum. Meth. A}{368}{1}{1995};
A.J. Buras and R. Fleischer, \Journal{\PLB}{365}{390}{1996};
N.G. Deshpande, X.G. He and S. Oh, in {\em hep-ph/9511462}.

\bibitem{Quigg}C. Quigg, private communication.

\bibitem{FerAih}F. Ferroni, these proceedings;
H. Aihara, these proceedings.

\bibitem{Ruc}W. Ruckstuhl, these proceedings.

\bibitem{But95}J.N. Butler ``Hadron Collider B Factories'',
Fermilab report LISHEP-95, unpublished.

\bibitem{Rij}M. Rijssenbeek, these proceedings.

\bibitem{Gro93}M. Gronau, \Journal{\PLB}{300}{163}{1993}.

\bibitem{Gro90}M. Gronau and D. London, \Journal{\PRL}{65}{3381}{1990}.

\bibitem{Des95}N.G. Deshpande and X.G. He, \Journal{\PRL}{74}{26}{1995}.

\bibitem{Sil94}J.P. Silva and L. Wolfenstein,
\Journal{\PRD}{49}{R1151}{1994}.

\bibitem{CLEO}CLEO Collab., \Journal{\PRL}{71}{3922}{1993}.

\bibitem{Gro96}M. Gronau and J. Rosner, \Journal{\PRL}{76}{1200}{1996};
see also A.S. Dighe, M. Gronau and J. Rosner in {\em hep-ph/9604233}.

\bibitem{Gro91}M. Gronau and D. Wyler, \Journal{\PLB}{265}{172}{1991}.

\bibitem{Dun91}I. Dunietz, \Journal{\PLB}{270}{75}{1991}.

\bibitem{LHCB}LHC-B Collab. Letter of intent, CERN preprint CERN/LHCC 95-5.

\bibitem{Ale92}R. Aleksan, I. Dunietz and B. Kayser, 
\Journal{\ZPC}{54}{653}{1992}.

\bibitem{Groetal}M. Gronau {\it et. al}, \Journal{\PRL}{73}{21}{1994};
and \Journal{\PRD}{50}{4529}{1994}.

\bibitem{Coyle}P. Coyle, these proceedings.

\bibitem{Ben96}M. Beneke, G. Buchalla and I Dunietz,
in {\em hep-ph/9605259}.

\bibitem{Dun95}I. Dunietz \Journal{\PRD}{52}{3048}{1995}.

\end{thebibliography}
\end{document}